\documentclass[preprint, aps, pre, preprintnumbers,amsmath,amssymb,showpacs]{revtex4}

 \usepackage[final]{graphicx}  

\newcounter{saveeqn}

\begin{document}

 \title{Competition between Traveling Fluid Waves of Left and Right Spiral Vortices and 
 Their Different Amplitude Combinations}

 \author{A.~Pinter, M.~L\"ucke, and Ch.~Hoffmann}
 \affiliation{Institut f\"ur Theoretische Physik, Universit\"at des Saarlandes,
 Postfach 151150, \\ D-66041 Saarbr\"ucken, Germany}

 \date{\today}

 \begin{abstract}
Stability, bifurcation properties, and the spatiotemporal behavior
of different nonlinear combination structures of spiral vortices in
the counter rotating Taylor-Couette system are investigated by full
numerical simulations and by coupled amplitude equation
approximations. Stable cross-spiral structures with continuously
varying content of left and right spiral modes are found. They
provide a stability transferring connection between the initially
stable, axially counter propagating wave states of pure spirals and
the axially standing waves of so-called ribbons that become stable
slightly further away from onset of vortex flow.
\end{abstract}


 \maketitle

The combination of two counter propagating plane linear traveling
waves (TWs) of equal amplitude creates a standing wave (SW). This
property is also the reason for the possibility of SW structures in
a variety of nonlinear pattern forming systems that show an
oscillatory instability and inversion symmetry in one or more
spatial directions \cite{CH93}. Then two symmetry degenerate counter
propagating TW structures as well as a nonlinear SW solution appear
at the common oscillatory bifurcation threshold. However, in many systems like,
e.g., binary mixture convection TW solutions bifurcate subcritically being 
initially amplitude unstable and then become strongly nonlinear on the stable
upper solution branch. Moreover, the SW solution is often amplitude unstable 
and/or unstable against phase propagation and TW formation.

Thus, we consider here nonlinear combination states of left-spiral
vortex (L-SPI) and right-spiral vortex (R-SPI) structures that ---
being mirror images of each other --- travel either upwards or
downwards in the annulus between two counter rotating cylinders of a
Taylor-Couette setup \cite{CI94, T94}. This system offers an easy
experimental and numerical access ({\em i}) to forwards bifurcating
initially stable TWs (i.e., L-SPI and R-SPI), ({\em ii}) to forwards
bifurcating SWs [called ribbons (RIBs) in the Taylor-Couette
literature \cite{CI94, T94}] that become stable slightly above
threshold, and in particular also ({\em iii}) to stable
cross-spirals (CR-SPI), i.e., combinations of L-SPI and R-SPI with
different amplitudes which provide a stability transferring
connection between TW and SW solution branches. This bifurcation
scenario does not seem to have been reported so far \cite{RIBs}.

Here, we describe stability, bifurcation properties, and the spatiotemporal
behavior of the three oscillatory structures ({\em i})-({\em iii}) by two
amplitude equations with a coupling that contains necessarily also a
quintic order term to enable CR-SPI solutions. Their results are compared with
numerical solutions of the full Navier-Stokes equations that we obtained for a
system with radius ratio $\eta=1/2$ by methods described in \cite{HLP04}.

The vortex structures ({\em i})-({\em iii}) are axially and
azimuthally periodic with wave numbers $k=2\pi /\lambda$ and $M$, respectively,
with $\lambda=1.3$ and $M=2$ throughout this paper.
They rotate with characteristic constant angular velocities into the same
direction as the inner cylinder \cite{HLP04} and thereby they are forced to
propagate axially except for RIB vortices which only rotate but do not propagate.

\begin{figure}
\includegraphics[clip=true,width=7.5cm, angle=0]{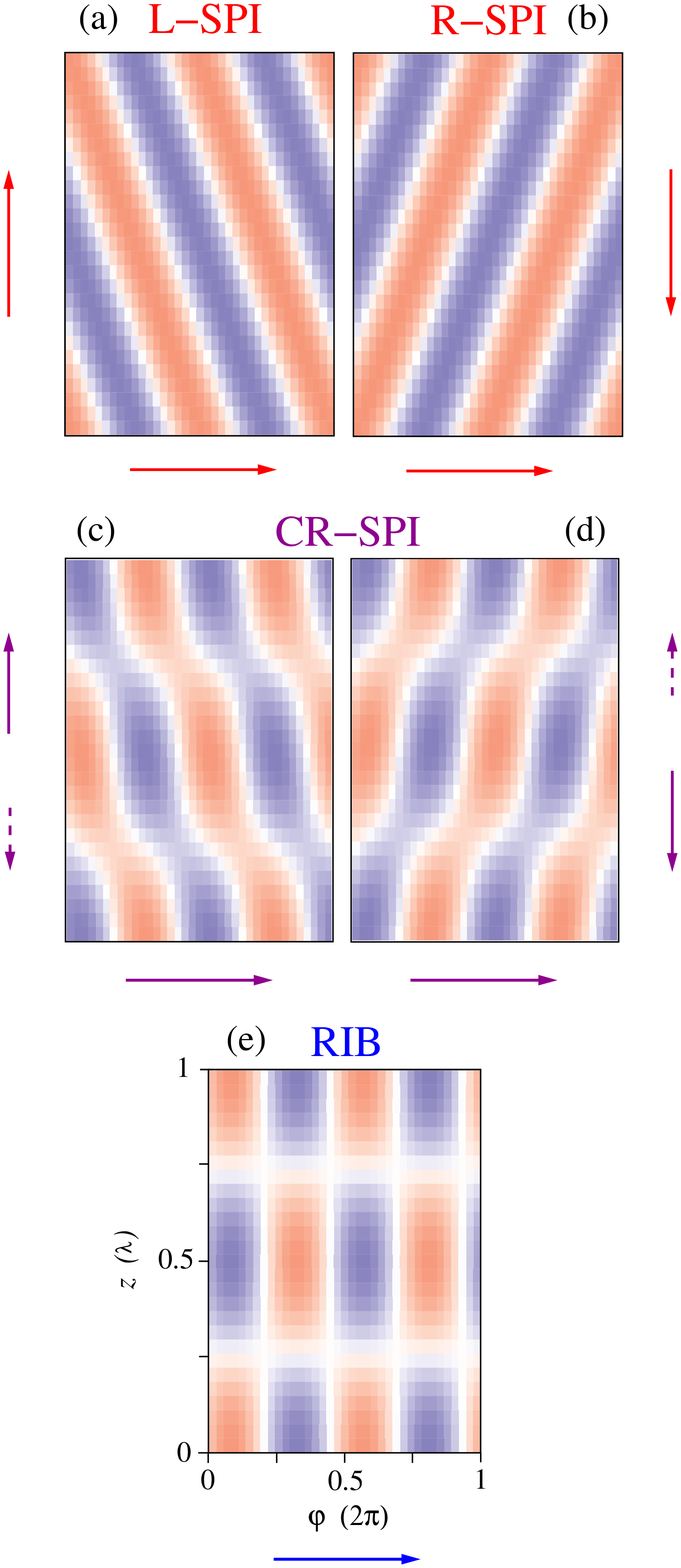}
\caption{(Color online) Snapshots of the radial flow $u$ in the
$\varphi-z$-plane of the 'unrolled' mid-gap cylindrical surface.
White identifies the nodes. Light (dark) grey [red (blue) in the electronic
version] marks radial out-flow (in-flow). Arrows indicates how the phase moves. For the CR-SPI in
(c) [(d)] the undulated pattern moves upwards [downwards] (full
arrows) while the undulation itself propagates slowly downwards
[upwards] (dashed arrows). Parameters are $R_2=-605.2 (a, b),
-604.725 (c,d), -604.5 (e)$ and $R_1=240$. \label{v-w-Plots}}
\end{figure}

In order to characterize the spatiotemporal properties of the
vortices we used also Fourier decompositions in azimuthal and axial
direction
\begin{eqnarray} \label{modenansatz}
f(r,\varphi,z,t) = \sum_{m,n} f_{m,n}(r,t)\,e^{i(m\varphi + nkz)}\,.
\end{eqnarray}
For example, L-SPI (R-SPI) flow contains only modes with $m=nM$
($m=-nM$) since it does not depend on $\varphi,z,t$ separately but
only on the phase combination $\phi_A= M\varphi + kz-\omega_A t$
($\phi_B= M\varphi - kz-\omega_B t$). Here
we use  $A$ and $B$ to identify L-SPI and R-SPI properties,
respectively. For SPI that are axial mirror images of each other the
modes in (\ref{modenansatz}) oscillate according to $f_{m,n}(r,t)
=\delta_{m,\pm nM} e^{-i(m/M)\omega t} \tilde{f}_{\pm n}(r)$ with a
common frequency $\omega_A= \omega_B=\omega(k,M)$. Thus, the
color-coded plots in Fig.~\ref{v-w-Plots}(a,b) of the SPI in the
$\varphi-z$ plane of an 'unrolled' cylindrical surface are patterns
of straight stripes. They both move in positive $\varphi$-direction,
i.e., into the direction of rotation of the inner cylinder
\cite{HLP04} with angular phase velocity $\dot{\varphi}_{SPI} =
\omega/M$. The L-SPI propagates upwards with axial phase velocity
$w_{ph}^A=\omega/k$ and and its mirror image the R-SPI propagates
downwards, $w_{ph}^B=-\omega/k$ .

On the other hand, a CR-SPI is a kind of combination of a L-SPI and
a R-SPI with different flow amplitudes and different frequencies.
Thus, the CR-SPI spectrum contains two frequencies, say, $\omega_S$
and $\omega_D$, that are in general incommensurate. Within the
amplitude equation approximation presented below $\omega_S=(\omega_A
+ \omega_B)/2$ and $\omega_D=(\omega_A - \omega_B)/2$. For the
CR-SPI in Fig.~\ref{v-w-Plots}(c) the amplitude of the L-SPI content
is larger than the R-SPI content while $\omega_A < \omega_B$. 
Then the periodic deformation of the L-SPI-dominated stripe pattern in
Fig.~\ref{v-w-Plots}(c) that is caused by the minority R-admixture
propagates axially downwards with velocity $\omega_D/k$ while 
rotating with angular velocity $\omega_S/M$.
The CR-SPI in Fig.~\ref{v-w-Plots}(d) is the mirror
image of Fig.~\ref{v-w-Plots}(c) with reverse amplitude and
frequency relations. For all CR-SPI investigated here
$\omega_S\gg|\omega_D|$. RIB states are degenerate CR-SPI states
with equal content of R-SPI and L-SPI contributions and $\omega_D=
0$ which do not propagate axially.

\begin{figure}
\includegraphics[clip=true,width=8cm, angle=0]{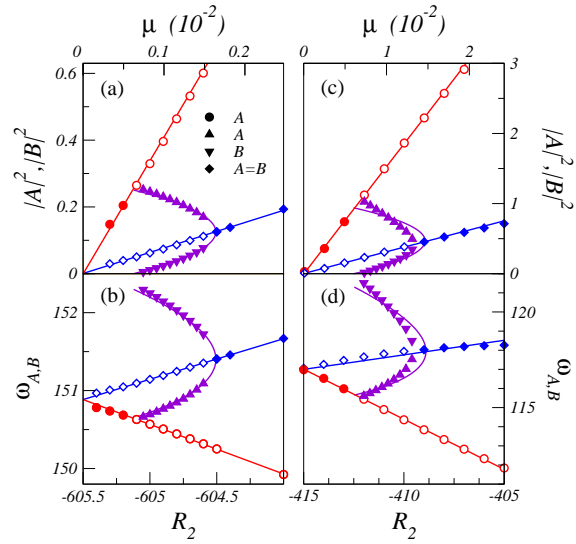}
\caption{(Color online) Bifurcation diagrams of SPI (red circles),
RIB (blue diamonds), and CR-SPI (purple triangles) obtained from numerical
solutions of the full Navier-Stokes equations for $R_1=240$ (a,b) and
$R_1=200$ (c,d) as functions of $R_2$ and $\mu$. Shown are the
squared mode amplitudes $|A|^2,|B|^2$ and frequencies $\omega_A,\omega_B$ of the
dominant modes $u_{2,1}(t),u_{2,-1}(t)$ (\ref{EQ:u2pm1}), respectively, in the
expansion (\ref{modenansatz}) of the radial velocity field $u$ at mid-gap.
Full (open) symbols denote stable (unstable) solutions \cite{control}. Lines
show amplitude-equation approximations. Here a CR-SPI with
$|A|>|B|, \omega_A<\omega_B$ transfers stability from an L-SPI ($A\neq 0, B=0$)
to the RIB state ($A=B$).
The analogous diagram with an R-SPI is obtained by interchanging $A$ and $B$.
\label{bifurcationdiagram}}
\end{figure}

Fig.~\ref{bifurcationdiagram} shows for two different $R_1$ bifurcation diagrams
of SPI, CR-SPI, and RIB solutions versus $R_2$ and its reduced distance
$\mu=(R_2 -R_2^0)/|R_2^0|$
from the onset, $R_2^0$, of SPI and RIB flow. Here $R_1$ and $R_2$ are the 
Reynolds numbers defined by the rotational velocities of the inner and outer
cylinder, respectively. Order
parameters are the squared amplitudes ($|A|^2,|B|^2$) and frequencies
($\omega_A,\omega_B$) of the dominant modes
\begin{equation} \label{EQ:u2pm1}
u_{2,1}(t)=|A|e^{-i\omega_A t} \quad , \quad u_{2,-1}(t)=|B|e^{-i\omega_B t}\,
\end{equation}
in the decomposition (\ref{modenansatz}) of the radial velocity $u$
at mid-gap. For all vortex structures investigated here the moduli
and frequencies of $u_{2,\pm1}$ in Eq.~(\ref{EQ:u2pm1}) are
constant.

The L-SPI ($A\neq 0, B=0$) shown in Fig.~\ref{bifurcationdiagram} by circles and
the RIB solution ($A =B$) marked by diamonds bifurcate at $\mu = 0$ with common
linear frequency $\omega^0$ out of the
unstructured CCF. However, with increasing $\mu$ their frequencies
vary differently. The squared amplitudes of these two states grow
basically linearly with $\mu$, albeit with different slopes.

Initially, the SPI is stable and the RIB is unstable. But then there appears
at larger $\mu$ a stable CR-SPI solution (triangles) which
transfers stability from the SPI to the RIB. The CR-SPI ($A \neq B$) bifurcates
with $B=0$ and finite $\omega_B$ out of the L-SPI as shown in
Fig.~\ref{bifurcationdiagram} (bifurcation out of
the symmetry degenerate R-SPI is analogous --- just exchange  $A$ and $B$). Then
$|B|$ grows and $|A|$ decreases while $\omega_A$ increases and $\omega_B$ becomes
smaller until the CR-SPI solution branches end with
$A=B$ and $\omega_A = \omega_B$ in the RIB state. The latter loses stability
outside the plot range of Fig.~\ref{bifurcationdiagram} to amplitude-modulated
CR-SPI that are not discussed here.

Spatiotemporal properties, bifurcation, and stability behavior of
the SPI, RIB, and CR-SPI states close to onset can reasonably well
be explained and described within an amplitude-equation approach.
Therein the fields are represented by the superposition
--- here written down, e.g., for the radial velocity $u(r,\varphi,z,t)$ ---
\begin{equation} \label{EQ:u-amp-repres}
u=\left[ A(t)\hat{u}_A(r)e^{ikz} + B(t)\hat{u}_B(r)e^{-ikz} \right]
e^{i(M\varphi-\omega^0t)} + c.c.
\end{equation}
of just the linear critical modes
$\hat{u}_{A,B}(r)e^{\pm ikz}e^{i(M\varphi-\omega^0t)}$ of L- and R-SPI with
slowly varying
amplitudes $A$ and $B$, respectively. Here we normalize $|\hat{u}_A|$ and
$|\hat{u}_B|$ to 1 at mid-gap. There the representation (\ref{EQ:u-amp-repres})
is quite sufficient. But closer to the
inner cylinder the vortex fields contain further axial and azimuthal modes that
are generated via nonlinear interactions of the critical modes. However,
it seems that close to onset they do not influence decisively the bifurcation
and stability behavior that is governed by the modes retained in the
approximation (\ref{EQ:u-amp-repres}).

In order to reproduce the bifurcation and stability behavior of
the aforementioned vortex states including the CR-SPI one needs
coupled equations for $A$ and $B$ of at least quintic order \cite{FDT91}.
Higher-order terms that are suggested in \cite{CI94} are not necessary to
ensure existence of CR-SPI solutions. Symmetry
arguments \cite{CI94} and simplicity considerations discussed below restrict the
form of the equations to
\begin{subequations} \label{EQ:amp-eq}
\begin{eqnarray}
\dot{A}=\left[ a\mu +b|A|^2 +c|B|^2 +  e\left(|A|^2 - |B|^2\right)|B|^2 \right] A \\
\dot{B}=\left[ a\mu +b|B|^2 +c|A|^2 +  e\left(|B|^2 - |A|^2\right)|A|^2 \right] B \, .
\end{eqnarray}
\end{subequations}
The coefficients $a-e$ are complex but only the moduli $|A|$ and $|B|$
enter into the square brackets. We have determined all coefficients
by comparing with bifurcation diagrams obtained for the flow
at mid-gap position.

We are seeking solutions of (\ref{EQ:amp-eq}) of the form
$A(t)=|A|e^{-i\Omega_A t} \, , \, B(t)= |B|e^{-i\Omega_B t}$
with constant moduli and frequencies. Here
$\Omega_{A,B}= \omega_{A,B} -\omega^0$ are the deviations of the
frequencies from the critical ones according to
(\ref{EQ:u2pm1}, \ref{EQ:u-amp-repres}).

In view of
Fig.~\ref{bifurcationdiagram} we have neglected here for simplicity reasons
quintic
contributions to SPI amplitudes and frequencies by discarding terms of the form
$d|A|^4A$ and $d|B|^4B$ in (\ref{EQ:amp-eq}a) and (\ref{EQ:amp-eq}b),
respectively. Taking
$a= 1 + a_i$ one then has $|A|^2=\mu/(-b_r)$ and $\Omega_A=-a_i\mu -b_i |A|^2 $
for, say, a L-SPI (for a R-SPI replace $A$ by $B$). The indices $r$ and $i$
denote real and imaginary parts,
respectively. As an aside we mention that a linear variation of $\Omega_A$
with $|A|^2$ holds for SPI also significantly further away from onset than in
Fig.~\ref{bifurcationdiagram}.

Motivated by Fig.~\ref{bifurcationdiagram} we also discard quintic contributions
to the modulus and to the frequency of the RIB state. This is enforced by
making in the terms $f|B|^4A$ and $f|A|^4B$ that appear on general grounds in
(\ref{EQ:amp-eq}a) and (\ref{EQ:amp-eq}b), respectively, the special choice
$f=-e$. Then the RIB solution is characterized by
$|A|^2=|B|^2=\mu/(-b_r-c_r)$ and $\Omega_A= \Omega_B= -a_i\mu -(b_i+c_i)|A|^2$.

The CR-SPI solution with finite $A \neq B$ exists only for nonzero
coupling $e$. It is most conveniently found and expressed in terms
of the combined order parameters
\begin{subequations}  \label{EQ:S-D-def}
\begin{eqnarray}
S = (|A|^2 + |B|^2)/2 \quad &,& \quad D = (|A|^2 - |B|^2)/2  \\
\Omega_S= (\Omega_A +\Omega_B)/2 \quad &,& \quad \Omega_D= (\Omega_A -\Omega_B)/2 \, .
\end{eqnarray}
\end{subequations}
Then the radial velocity field (\ref{EQ:u-amp-repres}) of the CR-SPI  at mid-gap
can be written into the form
$ u=\left[ |A|e^{i(kz-\Omega_Dt)} + |B|e^{-i(kz-\Omega_Dt)} \right]
e^{i(M\varphi-\omega^0t-\Omega_St)} + c.c.$. It
shows the aforementioned rotation frequency $\omega_S=\omega^0+\Omega_S$
and the axial propagation of the deformation that is governed by the
frequency $\omega_D=\Omega_D$.

The CR-SPI solution of (\ref{EQ:amp-eq}) reads
\begin{subequations} \label{EQ:CR-SPI-solution}
\begin{eqnarray}
S=(c_r-b_r)/(2e_r)\, &,& \,  D^2 = [\mu + (b_r+c_r)S]/(2e_r)\\
-\Omega_S=a_i\mu + (b_i+c_i)S -2e_i D^2 \, &,& \,
-\Omega_D=(b_i-c_i +2e_i S)D \, .
\end{eqnarray}
\end{subequations}
Note that $S$ is constant in this CR-SPI. $D$ grows via a pitchfork bifurcation
at $\mu_2=-(b_r+c_r)S$ out of the RIB ($D=0$) and connects at
$\mu_1= -2 b_r S$ with $D=|A|^2/2=S$ with the L-SPI and with $D=-|B|^2/2=-S$
with the R-SPI.

\begin{figure}
\includegraphics[clip=true,width=8cm, angle=0]{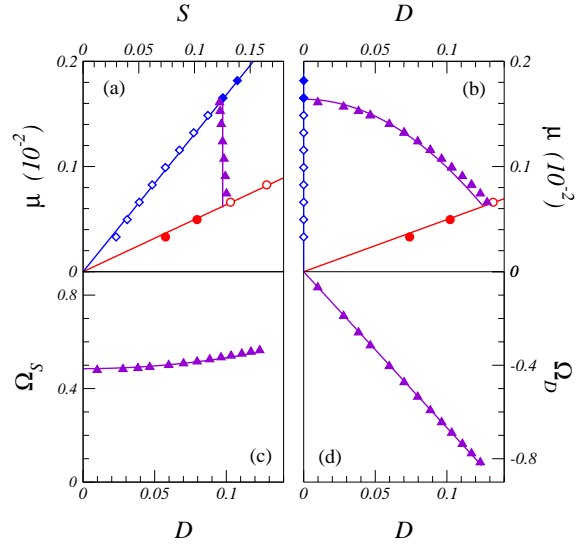}
\caption{(Color online)  Bifurcation diagrams of SPI (red circles),
RIB (blue diamonds), and CR-SPI (purple triangles) obtained from numerical
solutions of the full Navier-Stokes equations for $R_1=240$ using the order
parameters (\ref{EQ:S-D-def}): Relation between $\mu$ and
$S$ (a) and $D$ (b), respectively; CR-SPI frequencies
$\Omega_S=(\Omega_A+\Omega_B)/2=(\omega_A+\omega_B)/2-\omega^0$ (c) and
$\Omega_D=(\Omega_A-\Omega_B)/2=(\omega_A-\omega_B)/2=\omega_D$ (d),
respectively, versus $D$. Full (open) symbols  denote stable (unstable)
solutions \cite{control}. Lines show amplitude-equation approximations. Here a
CR-SPI ($|A|>|B|,D>0$ and $\Omega_A<\Omega_B,\Omega_D<0$)
transfers stability from an L-SPI ($B=0$, i.e., $D=S$) to the RIB state
($A=B$, i.e., $D=0$).
\label{FIG-R1=240}}
\end{figure}

\begin{figure}
\includegraphics[clip=true,width=8cm, angle=0]{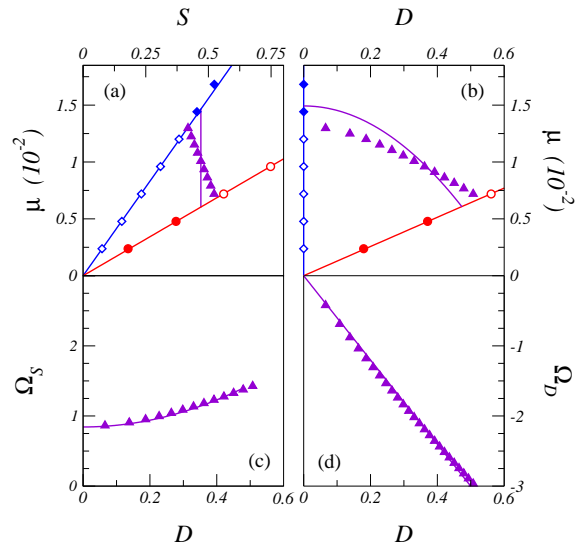}
\caption{(Color online) Bifurcation diagrams for $R_1=200$ as in
Fig.~\ref{FIG-R1=240}.
\label{FIG-R1=200}}
\end{figure}

\begin{figure}
\includegraphics[clip=true,width=7cm, angle=0]{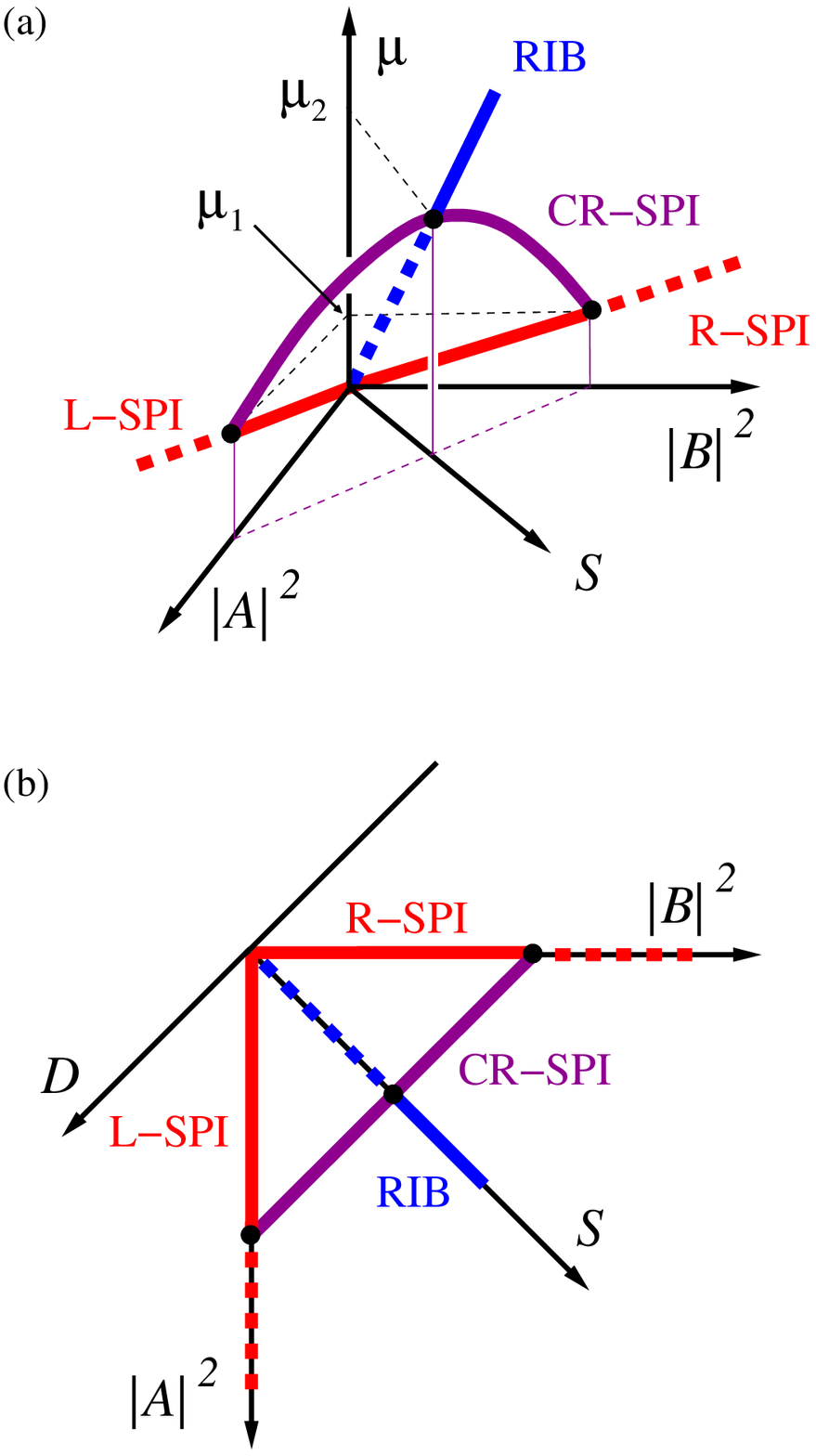}
\caption{(Color online)
Bifurcation diagrams of SPI, RIB, and CR-SPI resulting from the
amplitude-equation approximation. (a) 3D plot in $|A|^2-|B|^2-\mu$-space;
(b) projection onto the $|A|^2-|B|^2$-plane or the $D-S$-plane, respectively.
Full (dashed) lines denote stable (unstable) solutions. The CR-SPI
connects with constant $S$ the two SPI solution branches to the RIB thereby
transferring stability from the former to the latter.
\label{3d-Plot}}
\end{figure}

In Figs.~\ref{bifurcationdiagram} - \ref{FIG-R1=200} we compare
bifurcation diagrams of SPI, RIB, and CR-SPI solutions resulting from the
amplitude equations \cite{coefficients} (thin lines) with those from the full
numerical simulations (symbols). For the sake of presentational clarity the
stability of the former is not indicated in these figures. But
stability analyses of the three vortex states in question confirms that the
CR-SPI solution (\ref{EQ:CR-SPI-solution}) transfers stability between the RIB
and the SPI states as indicated in Fig.~\ref{3d-Plot}.

We should like to mention that keeping the discarded quintic terms by setting
$d\neq 0$ and $f\neq -e$ slightly
improves the agreement between the solutions of the Navier-Stokes equations and
the amplitude equation approximation without changing the bifurcation topology
and the stability behavior of the latter. However, then the CR-SPI solution
would still have $S=$ const. To
obtain deviations from this behavior as displayed, e.g., by the triangles
of Fig.~\ref{FIG-R1=200}(a) one needs higher orders in eqs.~(\ref{EQ:amp-eq}).

So, the model (\ref{EQ:amp-eq}) is a minimal one in
the sense that the quintic terms that are necessary to ensure the existence of
$A \neq B$ solutions enter only into the coupling between R- and L-SPI
modes but not into the SPI and RIB solutions.
Note furthermore that the coupled amplitude equations (\ref{EQ:amp-eq}) apply
equally well to other systems with two distinct $A$ and $B$ solutions --- be
they stationary or oscillatory --- that
are related to each other by analogous (symmetry) requirements with similar
nonlinear couplings. Thus, bifurcation diagrams in which the pure
$A$, the pure $B$, and the $A=B$ solutions are connected with a stability
transfer by a $A \neq B$ solution branch could also arise in these systems with
a form that would be (topologically) similar to that of Fig.~\ref{3d-Plot}.

This work was supported by the Deutsche Forschungsgemeinschaft.

\newpage
 
\clearpage

\begin{thebibliography}{999}
\bibitem{RIBs}
Stable RIBs were found \cite{TESM89} in a long system of
aspect ratio $\Gamma=36$ with radius ratio $\eta=0.727$ where theory
\cite{DI84,CI94} and numerical simulations \cite{TESM89} predicted a subcritical
transition. However, no CR-SPI were obseved. RIBs in end-plate-dominated short
systems ($\Gamma < 10, \eta=1/2$) were recently reported  \cite{LPA03-04} to come
with two different symmetries \cite{KP92}.
\bibitem{coefficients}
We have determined the coefficients as follows: a linear
analysis of the flow dynamics gave $\omega^0= 150.9 \,\,(117.0)$ and
$a_i= 45.13\,\,(15.97)$ for $R_1=240$ ($R_1=200$).
Then $b_r$ and $b_i$ were obtained by fitting $|A|^2$ and
$\Omega_A$, respectively, of the nonlinear numerical SPI solution. Thereafter
$c_r$ and $c_i$ were fixed analogously with the numerical RIB
solutions. Finally, $e$ was determined with the CR-SPI.
The results are
\mbox{$100 b=-0.245+i\,83.54\,\,(-0.659+i\,132.8)$},
\mbox{$100 c=-1.06-i\,554\,\,(-2.517-i\,383.9)$},
\mbox{$100 e=-3.251+i\,118.4\,\,(-1.976+i\,89.2)$}.
\bibitem{control}
Unstable SPI solutions were obtained by suppressing the $m=0-$mode of $u$. This
stabilizes the SPI against the transition to CR-SPI and
RIB states which have finite $m=0-$ content. The unstable RIB solution was
stabilized by imposing  the $z \rightarrow -z$ symmetry of this state thereby
suppressing SPI and CR-SPI solutions.
\end{thebibliography}
\end{document}